\documentstyle[amstex,amssymb,twocolumn,aps,prl,epsfig]{revtex}

\begin{document}
\title{Is there a maximum observable redshift in an open universe?}
\author{Julio A. Gonzalo}
\address{Facultad de Ciencias \\
Universidad Aut\'{o}noma de Madrid.\\
Cantoblanco, 28049 Madrid\\
Spain}
\maketitle

\begin{abstract}
An estimate of the maximum observable redshift is obtained using only $%
t_{0}\cong (14\pm 3)\times 10^{9}years$, $H_{0}\cong 65\pm 10Km\sec
^{-1}Mpc^{-1}(t_{0}H_{0}\cong 0.91\pm 0.08/0.18)$assuming $\Lambda \cong 0.$%
The resulting $\max $imum redshift $z_{+}\cong 10$ appears to give a
reasonable upper limit to the highest actually observed redshifts. Some
implications are discussed.
\end{abstract}


As pointed out about one quarter of a century ago by Beatriz M. Tinsley\cite
{B.M.}, the dimensionless product $H_{0}t_{0}$ (Hubble's constant times the
time elapsed since the big bang) could turn out to be the most tractable
method to elucidate the actual large scale cosmic dynamics, rather than
using the dimensionless ratio $\Omega _{0}=\frac{\rho _{0}}{\rho _{c}}$,
where $\rho _{c}=\frac{3H_{0}^{2}}{8\pi G}.$She pointed out that, although
the problems to find well defined distance scales to get $H_{0}$, and to
estimate the ages of the oldest stars to get $t_{0}$, were (and still are)
quite formidable, they were probably not as difficult as those encountered
to determine small redshift differences for nearby and for distant galaxies.
Her warning \cite{B.M.}that the test of $(H_{0}t_{0})$ ''should be kept in
mind when estimates of $H_{0}$ and $t_{0}$ are refined'' should be
remembered now that we have improved observational data \cite{Physics}. The
test requires only local data, as she noted, which implies a definite
advantage.

It is well known that at very early times Einstein's cosmological equations
for k%
\mbox{$<$}%
0 (open), k=0 (flat) and k%
\mbox{$>$}%
0 (closed) universes are alike in predicting $Ht\cong 2/3$ and $\Omega \cong
1.$ This is true for a zero and a vanishingly small cosmological constant, $%
\left| \frac{\Lambda }{3H^{2}}\right| <<1$. In this work we will use refined
recent values \cite{Astro} for $H_{0}=\stackrel{\cdot }{R}_{0}/R_{0}\cong
65\pm 10Km\sec ^{-1}Mpc^{-1}$ and $t_{0}\cong (15\pm 2)\times 10^{9}years$
to specify global cosmic dynamics from the open solutions of Einstein's
equations.

The total mass density can be given by $\rho =\rho _{m}+\rho _{\gamma }+\rho
_{\Lambda },$ where $\rho _{m}$ is the matter mass density, $\rho _{\gamma }$
the radiation mass density (now $\rho _{\gamma 0}<<\rho _{m0}$) and $\rho
_{\Lambda }=\Lambda /8\pi G>0$ (moderately repulsive) is the mass density
associated with small but nonvanishing cosmological constant $\Lambda .$

For $k<0$ (open universe) and $\Lambda =0$ Einstein's cosmological equation
can be written as

\begin{gather}
\stackrel{\cdot }{R}=R^{-1/2}\left\{ \frac{8\pi G}{3}\rho R^{3}+\left|
k\right| c^{2}R\right\} ^{1/2}=  \label{eq1} \\
=R^{-1/2}\left\{ \frac{8\pi G}{3}\rho _{+}R_{+}^{3}+\left| k\right|
c^{2}R\right\} ^{1/2}  \nonumber
\end{gather}
where $R_{+}$is defined precisely at $\frac{8\pi G}{3}\rho
_{+}R_{+}^{3}=\left| k\right| c^{2}R_{+}$. Then, for $R<<R_{+},$ the term
involving $\left| k\right| $ within the square root in \ref{eq1} becomes
negligible, i.e. we are in the early (Einstein-de Sitter) phase of the
expansion. For $R\gtrsim R_{+},$on the other hand, both terms within the
square root become important, and we enter into the explicitly open phase of
the expansion.

The parametric solutions of \ref{eq1} can be given \cite{Noe}as a function
of $R_{+},$ which has the meaning of a Schwarzschild radius for the total
mass $M_{+}=\frac{4\pi }{3}\rho _{+}R_{+}^{3}=\frac{4\pi }{3}\rho
_{0}R_{0}^{3}$ of the universe, assumed to be conserved. They are given by 
\begin{equation}
t=\frac{R_{+}}{\left| k\right| ^{1/2}c}(\sinh y\cosh y-y),\text{ \ \ }%
R=R_{+}\sinh ^{2}y  \label{eq2}
\end{equation}
It is easy to check from \ref{eq2} that the dimensionless product $Ht=(%
\stackrel{\cdot }{R}/R)t$ is given by 
\begin{equation}
Ht\equiv \frac{\left\{ \sinh y\cosh y-y\right\} \cosh y}{\sinh ^{3}y}\geq 
\frac{2}{3}  \label{eq3}
\end{equation}

Present observational constraints\cite{Astro} on $H_{0}$ and $t_{0}$ result
in $H_{0}t_{0}\cong 0.91>2/3$, with 
\begin{equation}
(H_{0}t_{0})_{\min }\geq 0.73>2/3  \label{eq4}
\end{equation}
and 
\begin{equation}
\Omega _{0}\cong 0.2\pm 0.1,\text{ \ \ \ \ \ \ \ \ \ \ \ \ \ }\Omega
_{0}=\Omega _{0m}+\Omega _{0\gamma }+\Omega _{0\Lambda }  \label{eq5}
\end{equation}

The data on $H_{0}(65Km/sMpc)$ and $t_{0}(13.7\times 10^{9}years)$ which
imply $H_{0}t_{0}\cong 0.91$, are sufficient, together with \ref{eq3}, to
specify the redshift $z_{+}$ corresponding to $R=R_{+}$ (Swarzschild
radius), by means of 
\begin{equation}
y_{0}=\sinh ^{-1}(R_{0}/R_{+})^{1/2}\cong \sinh ^{-1}\left( \frac{1+z_{+}}{%
1+0}\right) \cong 1.92  \label{eq6}
\end{equation}
taking into account that the scale factor $R$ is related \cite{Peebles}to
the redshift by means of

\[
R=R_{0}/(1+z),\qquad \longrightarrow \qquad R_{+}=R_{0}/(1+z_{+}) 
\]

At $R<R_{+}$ no protogalaxies (no early quasars) could have become
gravitationally bound yet. Then one is entitled to expect that $%
z(R_{+})=10.1 $ (from \ref{eq6}) is the highest possible observable redshift
in our open universe. Estimates of galaxy formation epoch redshifts \cite
{Peebles} are not inconsistent with $z_{+}=z(R_{+})\cong 10$ as an upper
observable redshift limit.

After decoupling/atom formation ($\rho _{\gamma }=\rho _{m})$ taking place 
\cite{Noe}at $T_{af}\cong 3880K$ the universe became transparent. The data
on $H_{0},t_{0}$ together with data on $T_{0}$ and $\ T_{af}$ indicate that
radiation/matter mass density equality ($\rho _{\gamma }\cong \rho _{m})$
and atom formation ($\rho _{\gamma }(T_{af})$ with $T_{af}$ as given by
Saha's law \cite{Noe}) occur at the same cosmic time.

Table I gives the Hubble's constant, the density parameter and the redshift
corresponding to atom formation density ($\rho _{af})$, protogalactic
density ($\rho _{g})$, \ Schwarzschild $\left( \rho _{+}=\frac{M_{+}}{\frac{%
4\pi }{3}R_{+}^{3}}\right) $ and present density ($\rho _{0})$ determined at
the respective cosmic temperature. It may be noted that $\Omega _{0}\cong
0.082$ is below the generally estimated value $\Omega _{0}\cong 0.2\pm 0.1$,
but not very far. A non-zero $\Lambda $ may alter the effective values of $%
R_{+}$ and $T_{+}$, but we may hope that changes in $H_{+},\Omega _{+}$ and $%
z_{+}$ are not too drastic.

In summary, a direct estimate of the maximum observable redshift $z_{+}\cong
10$ is obtained for an open universe with negligible cosmological constant
which appears to be compatible with observed redshifts for the oldest
galaxies and quasars. It may be noted that recent evidence\cite{Perlmutter} 
\cite{Schmidt}in favor of an accelerated expansion of the universe from type
Ia supernovae is consistent \cite{Axenides} with $R_{\Lambda }=\left(
\Lambda \right) ^{-1/2}\cong 10^{28}cm$ which is somewhat larger but of the
order of $R_{+}\cong 2GM_{+}/c^{2}\cong 1.33\times 10^{27}cm.$ This might be
taken as an indirect indication that $z_{+}\cong 10$ may be not too
drastically affected by a $\Lambda $ of this order.

In any event our expectation of $(z_{obs})_{max}\simeq 10$ is ''testable''
by comparison with new incoming data from the more modern and powerful
telescopes.

We wish to thank Angeles Diaz Beltran, UAM, for reading the manuscript and
helpful criticism and A. Hewish, for encouragement and advice.

\onecolumn
{\small Table I. Redshifts at various cosmic events }

{\small $
\begin{array}{cccccc}
Event & T(K) & y=\sinh ^{-1}(T_{+}/T)^{1/2} & H(y)(\text{Km/secMpc}) & 
\Omega (y)(\text{dimensionless}) & z(\text{dimensionless}) \\ 
\text{Atom formation (}\rho _{m}=\rho _{\gamma }) & 3880 & 0.0884 & 
1.00\times 10^{6} & 0.9922 & 1422 \\ 
\text{Protogalaxies (}\rho \cong \rho _{g}) & 300 & 0.3131 & 2.26\times
10^{4} & 0.9080 & 109 \\ 
\text{Schwarzschild }\left( \rho _{+}=M_{+}/\frac{4\pi }{3}R_{+}^{3}\right)
& 30.4 & 0.8813 & 982 & 0.5000 & 10.1 \\ 
\text{Present (}\rho _{0}\cong 6.54\times 10^{-31}g/cm^{3}) & 2.726 & 1.9206
& 65 & 0.0823 & 0
\end{array}
$ }

{\small \twocolumn
}

\end{document}